\documentclass[twocolumn,superscriptaddress,prl,aps,floats,showpacs]{revtex4-1}
\usepackage{amsmath,amssymb,amsfonts,amsthm}
\usepackage{graphicx}
\usepackage{natbib}
\usepackage{bm}
\usepackage{color}

\begin{document}
\title{The Hausdorff Dimension of Two-Dimensional Quantum Gravity}
\author{Bertrand Duplantier}
\affiliation{Institut de Physique Th\'{e}orique, CEA/Saclay, F-91191
Gif-sur-Yvette Cedex, France}

\date{August 15, 2011}

\begin{abstract}
We argue that the Hausdorff dimension  of a quantum gravity random surface is always $D_{\mathcal H}= \bf 4$,  \textit{irrespective} of the conformal central charge $c$, $-2 \leq c \leq1$, of a critical statistical model possibly borne by it. The Knizhnik-Polyakov-Zamolodchikov (KPZ) relation  allows us to determine $D_\mathcal H$ from  the exact Hausdorff dimension 
of random maps with large faces, recently rigorously studied by Le Gall and Miermont, and reformulated by Borot, Bouttier and Guitter as a  loop model on a random quadrangulation. This contradicts several earlier conjectured formulae for $D_{\mathcal H}$ in terms of $c$. 
\end{abstract}
\maketitle
\textit{Introduction.}---Understanding and describing the quantum nature of gravity is one of the most critical issues in physics. Thirty years ago,  Polyakov introduced  \textit{Liouville quantum gravity} \cite{MR623209} 
to describe the random geometry of world sheets of gauge fields and strings. This seminal work  was followed by a flurry of studies of this novel ``quantum geometry'' \cite{MR1465433}.  Such 2D random surfaces are also expected to arise as continuum limits of the random-planar-graph surfaces developed in \textit{random matrix theory}, as became clear after  KPZ discovered their famous  relation   between critical exponents on a random surface and in the Euclidean plane \cite{MR947880,MR981529,*MR1005268,2009arXiv0901.0277D,*springerlink:10.1007/s00222-010-0308-1}. The  fractal properties of 2D gravity  are  thereby understood  from its deep relation to conformal field theory (CFT) \cite{[{See, e.g., }] [{, and references therein.}]2004IJMPA..19.2771N}.  A rigorous description in terms of the so-called Schramm-Loewner evolution (SLE) is also  being developed \cite{2010arXiv1012.4800D,*2010arXiv1012.4797S}. 
   
More recently, a combinatorial approach to this subject has emerged. It uses bijections from random planar graphs, or ``maps'', to labelled random trees. 
It allows a refined probabilistic description, in particular of  \textit{metric} or  \textit{geodesic} properties \cite{CS2004,*2002Angel,*1208.05135,Bouttier2003535} not yet accessible in Liouville or matrix approaches, and of universal scaling limits \cite{2011arXiv1101.4856L}.     
   A major effort now seeks to prove that all these approaches converge to a same universal continuous object,  possibly coupled to a  statistical system or CFT.
      
   One of the most enduring questions concerns the \textit{intrinsic Hausdorff dimension} $D_\mathcal H$, that describes how the area of a quantum ball, $\mathcal A \asymp {\mathfrak l}^{D_\mathcal H}$, scales with its geodesic radius $\mathfrak l$.  
   For the so-called \textit{pure gravity} case of large planar maps, one has  $D_\mathcal H =4$   
 \cite{Kawai199319,MR1338099} as combinatorially proven in \cite{CS2004,*2002Angel,*1208.05135}, or  in  \cite{Bouttier2003535} via the two-point function.
  
 When ``matter'' is present, that is, when the surface bears a critical system represented by a CFT of central charge $c \leq 1$, the situation is much less  clear. 
 A first formula, $D^{(1)}_{\mathcal H}=\frac{\sqrt{25-c}}{\sqrt{1-c}}-1$, was proposed  in the so-called ``string field theory'' framework \cite{1990IJMPA...5.1093D,*PhysRevD.50.7467}. An improper representation of the geodesic distance there was pinpointed in \cite{Bowick1997197}, and a value $D_\mathcal H (c=1/2)=4$ derived for the Ising model case. 
 Another formula, $D^{(2)}_{\mathcal H}=2\frac{\sqrt{25-c}+\sqrt{49-c}}{\sqrt{25-c}+\sqrt{1-c}}$, was then proposed  \cite{1993PThPS.114....1W}  and claimed to  fit well results from numerical simulations, in particular for $c\in [-2,0]$  \cite{PhysRevLett.68.2113,Ambj퓊n1998673,Kawamoto2002533}. However,  it has been consistently observed  that, at least for $c\geq 0$, a value $D_{\mathcal H} \approx 4$ is also plausible \cite{Ambj퓊n1995313,*Catterall199558}. 
 In this Letter, we argue that   one has $D_\mathcal H=4$ independent of the value of $c$, for  $c\in[-2,1]$. We focus  
 on the case of a random surface bearing a critical loop model, 
 and use recent combinatorial results  \cite{1204.05088,2011arXiv1106.0153B}. 
 
 Le Gall and Miermont  \cite{1204.05088}   
recently introduced a model of random planar maps with (macroscopically)  \textit{large faces} in the scaling limit.  They proved rigorously that these maps have, with probability one, a parameter dependent Hausdorff dimension 
 $D_\mathcal G$ continuously varying between $2$ and $4$.  As anticipated by them, and shown recently by Borot, Bouttier and Guitter \cite{2011arXiv1106.0153B}, these maps can also be obtained from maps  bearing  an $O(n)$ loop model. The so-called \textit{gasket}, defined by  erasing  the interiors of all the outermost loops, is such a planar map with large faces. In this work  we argue that the underlying dimension  $D_\mathcal H$ can be derived from  $D_\mathcal G$ by using the KPZ relation for the $O(n)$ model gasket.  
We first describe the essentials of the constructions in \cite{1204.05088,2011arXiv1106.0153B}.  
\begin{figure}[h]
\begin{center}
\includegraphics[angle=0,width=.63290\linewidth]{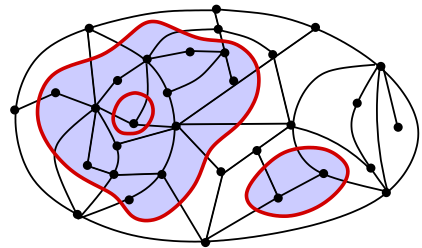}
\end{center}
\caption{\label{loopconfig}   (Courtesy of E. Guitter  \cite{2011arXiv1106.0153B}.)  A random planar quadrangulation with a boundary of length $2p$ ($=8$), and an $O(n)$ loop model on the dual lattice. The configuration here of $28$ quadrangles (quads) decomposes into  $\mathcal N_0=9$ empty quads, $\mathcal N_1=8$ trans-quads, and $\mathcal N_2=11$ cis-quads, building up 3 loops, and resulting in a weight  $\mathcal W_{O(n)}=n^3(h_0)^9 (h_1)^8 (h_2)^{11}$. The gasket $\mathcal G$  results from erasing the interior (shaded area) of all outermost loops, and retaining only those vertices which lay outside of the loops. This yields a bipartite planar map $\mathcal G$ with the same boundary, and made here of five of the original empty quads and of two larger faces of degrees $2k=6$ and $10$, with a total weight $W_{\bf q }({\mathcal G})=(q_2)^5(q_3)^1(q_5)^1$ \cite{1204.05088}.}
\end{figure}

\textit{Planar maps with large faces}.---A planar map  is a proper embedding of a finite connected graph $\mathcal G$ on the two-dimensional sphere, and is said to be bipartite when all its faces have even degrees (number of edges).  Given a sequence of nonnegative real numbers  ${\bf q }:=(q_k)_{k\in \mathbb N}$,  one defines  the Boltzmann weights  
$W_{\bf q }({\mathcal G}):=\prod_{f\in \mathfrak F({\mathcal G})}q_{\textrm{deg}(f)/2}$, 
where $\mathfrak F({\mathcal  G})$  is the set of faces of ${\mathcal G}$ and $\textrm{deg}(f)$ is the (even) degree of face $f$ \cite{1204.05088}. More precisely, one considers   sequences  of the form \cite{1204.05088} 
\begin{eqnarray}\label{q}
q_k= c_{\circ} \beta^{k} q_k^\circ,\,\,\,q_k^\circ \stackrel{k\to \infty}{\sim}k^{-a}, \,\,\, a\in (3/2,5/2),
\end{eqnarray}
where  $(q_k^{\circ})_{k\in \mathbb N}$ is a sequence of nonnegative real numbers with a \textit{heavy tail}, and  
$\sim$ means the ratio of the two sides has limit 1. The  constants $c_\circ$ and $\beta$ are fine tuned to this sequence, via the auxiliary series 
$f_\circ(z):=\sum_{k\geq1} {2k-1\choose k}q_k^\circ (\frac{z}{4})^{k-1}$. Using  Stirling's asymptotics 
 for the binomial coefficient,  
its radius of convergence is 1 with finite values $f_\circ(1)$ and $f_{\circ}'(1)$, because  $a>3/2$. One takes  $c_\circ:=4/f'_\circ(1)$ and $(4\beta)^{-1}:=1+f_\circ(1)/f'_\circ(1)$ \cite{1204.05088}. 

\textit{Scaling limit and Hausdorff dimension}.---For the particular choice \eqref{q} of face statistics with a heavy tail, 
the authors are able to prove rigorously that a \textit{scaling limit} of these random maps exists,  where ``macroscopic'' faces remain present.  A continuous so-called distance process can then be defined  \cite{1204.05088}, corresponding to a rescaling of the discrete graph distance by a factor $|{\mathcal G}|^{-1/2\alpha}$, with $\alpha:=a-1/2$, when the number of edges $|{\mathcal G}|$ of $\mathcal G$ tends to $\infty$.  
These random planar maps thus have an exact Hausdorff dimension 
\begin{equation}
\label{haus}
D_\mathcal G=2\alpha=2a-1 \in (2,4), 
\end{equation}
and are in a \textit{different universality class} from that of random planar maps with faces of \textit{bounded} or \textit{fast decaying degrees}, that of pure gravity of Hausdorff dimension $4$ \cite{CS2004,*2002Angel,*1208.05135}. 

\textit{Boundary partition function}.---As indicated in \cite{2011arXiv1106.0153B}, it is useful to consider maps with a boundary of fixed length $2p$ ($p\geq 1$),  called $p$-maps. They are  bipartite maps on the sphere, rooted along some (boundary) edge, whose external (empty) face has a fixed degree $2p$ (Fig.~\ref{loopconfig}). Their partition function ${\bf Z}^{(p)}_{\bf q }$ is defined as the total weight of all such $p$-maps $\mathcal G$, with individual weights $W_{\bf q }(\mathcal G)$ given by  \eqref{q}.  Asymptotically, as shown by combinatorial analysis \cite{2011arXiv1106.0153B},  the face weights \eqref{q}  are in fact  directly related to these  partition functions:   
\begin{equation} \label{zpqqk}
 q_k\stackrel{k\to \infty}{\sim} 2\sin \pi(a-3/2)\,  \beta^{2k} \,{\bf Z}^{(k)}_{\bf q },
  \end{equation} 
a key equivalence for the relation to an $O(n)$ loop model, to which we turn now.
 
 \textit{$O(n)$ model on quadrangulations}.---In this model \cite{2011arXiv1106.0153B}, 
 self- and mutually-avoiding loops are drawn on a tetravalent graph \textit{dual} of a quadrangulation, each loop carrying a non local weight $n$ (Fig.~\ref{loopconfig}). They delineate three types of quadrangles (quads): \textit{empty} ones not traversed by any  loop, of local weight $h_0$,  \textit{trans-quads} crossed by a loop  at opposite edges (local weight $h_1$), and \textit{cis-quads} with loop crossings at adjacent edges (local weight $h_2$) (Fig.~\ref{loopconfig}). A  quadrangulation carrying $\mathcal L$ loops, made of $\mathcal N_0$ empty quads, $\mathcal N_1$ trans-quads and $\mathcal N_2$ cis-quads,  receives a weight $\mathcal W_{O(n)}:= n^\mathcal L h_0^{\mathcal N_0} h_1^{\mathcal N_1} h_2^{\mathcal N_3}.$

A $p$-\textit{quadrangulation} with a boundary of length $2p$ ($p\geq 1$) is then defined as a  planar map whose external face has degree $2p$, all  other faces being quadrangles. The partition function $\mathcal Z^{(p)}_{O(n)}$   
 is then obtained by summing the weights  $\mathcal W_{O(n)}$ over all such (rooted) $p$-quadrangulations equipped with all possible collections of loops on the dual graph. 

For a given $n\in [0,2]$, there exists a critical surface $\mathcal C(n)$ in the parameter space $\bar{h}:=(h_0,h_1,h_2)\in \mathbb R_+^3$. It is encoded by a positive decreasing function $h_0^*(h_1,h_2,n)$ s.t.~the model is well-defined  in the bounded domain $0\leq h_0< h_0^*$, \textit{critical} on the surface $h_0=h_0^*$, ill-defined elsewhere. On $\mathcal C(n)$, there exists a  doubly critical line $\mathcal C_*(n): h_1=h_1^*(h_2,n)$, $h_0=h_0^*(h_1^*,\cdot,\cdot)$, where the model is in the so-called  \textit{dilute critical phase};  in the open sets $\mathcal C_{\pm}(n): h_0=h_0^*(h_1\gtrless h_1^*,\cdot,\cdot )$  the model is in the  \textit{dense critical phase} ($\mathcal C_+$) or \textit{pure gravity phase} ($\mathcal C_-$). This description of critical loci  
is corroborated by exact solutions of the $O(n)$ model on random trivalent graphs \cite{1989MPLA....4..217K,*1990NuPhB.340..491D}, as well as by that   
 of the model at hand in the $(h_1\geq 0,h_2=0, h_0\geq 0)$ sector  \cite{2011arXiv1106.0153B}.

In the scaling limit, the dense phase $\mathcal C_+$ is characterized  by macroscopic fractal loops that  bounce off themselves and off each other infinitely often; in the dilute phase $\mathcal C_*$, these loops have no self nor mutual intersections;  in the pure gravity phase $\mathcal C_-$, they stay microscopic,  vanishing in the scaling limit. (See below.) 
 
\textit{Asymptotics}.---On the critical surface $\bar h\in \mathcal C_+ \cup \mathcal C_*$, the so-called disk partition function has the distinctive asymptotic behavior  \cite{2011arXiv1106.0153B,Moore1991665,*Kostov1995284}
\begin{eqnarray}\label{Zonasymp}
\mathcal Z^{(p)}_{O(n)}({\bar h})\stackrel{p\to \infty}{\sim} c_\bullet\, \beta^{-p} p^{-a},\,\,\,
\beta=h_1+2h_2,
\end{eqnarray}
where $c_\bullet$ is a factor independent of $p$. The \textit{universal critical exponent} $a$ in \eqref{Zonasymp} is given by 
\begin{eqnarray}\label{na}
&n&=2\sin \pi (a-3/2) \in (0,2),\\ \nonumber
&a&\in (3/2,2),\,\, \bar h \in \mathcal C_+;\,\,\,
a\in (2,5/2),\,\,  \bar h \in \mathcal C_*,  
\end{eqnarray}
the two ranges of $a$ corresponding to the dense and dilute phases, respectively, and $a=2$  to the limit case $n=2$ where these  phases coincide. In the pure gravity phase  $\bar h\in \mathcal C_-$, one has $a=5/2$. 

\textit{Relation to bipartite maps with large faces}.---From \eqref{q}, \eqref{zpqqk} and \eqref{Zonasymp}, one observes that both partition functions share the same asymptotic  behavior.  
 This strongly suggests that the two models can in fact be put in \textit{bijection} one to another, this resulting in a complete identity of their partition functions  $\mathcal Z^{(p)}_{O(n)}={\bf Z}^{(p)}_{\bf q }, \forall p \geq 1$. This relation was introduced in   
\cite{1204.05088} and established combinatorially  in \cite{2011arXiv1106.0153B} for the present model.  

As can be observed geometrically,  faces of arbitrary degree $2k$  can be drawn as contours of outermost loops of length $2k$ in the $O(n)$ model  (Fig.~\ref{loopconfig}). 
When $k$ is large, the contour loop, of weight $n$, of a face of length $2k$ crosses a long \textit{thin quads' rim}  of width essentially \textit{one} and length $2k$.  The result is a weight $(h_1+2h_2)^{2k}$,   
each factor corresponding to one choice of trans-quad and two choices of cis-quads. Inside the rim lies a face of large degree of order $2k$ and  containing an $O(n)$ $2k$-quadrangulation of total weight  $\mathcal Z^{(k)}_{O(n)}$. The result is an overall face weight \cite{1204.05088,2011arXiv1106.0153B}:
 \begin{eqnarray}
\label{equiv1}
q_k\stackrel{k\to \infty}{\sim} n\, \beta^{2k} \mathcal Z^{(k)}_{O(n)}(\bar h),
\end{eqnarray}
 with $\beta=h_1+2h_2$.   Then \eqref{equiv1} exactly coincides with  \eqref{zpqqk} in the face model, in agreement with  the  relation \eqref{na} \cite{2011arXiv1106.0153B}. Lastly, comparing \eqref{q},   \eqref{Zonasymp}  and \eqref{equiv1}  
 gives $c_\circ=n\,c_\bullet$. 
 
 To summarize, 
 erasing the content of the outermost loops of an $O(n)$ model defines its gasket $\mathcal G$. For a critical model  on a random quadrangulation, parameterized as in \eqref{na}, this yields the large face planar map model with  weights   \eqref{q}. \textit{Therefore, the gasket $\mathcal G$  of a critical $O(n)$ model on a random surface $\mathcal S$,  defined as  the scaling limit of the random quadrangulation, has a Hausdorff dimension rigorously given by \eqref{haus}.} 

\textit{Coulomb gas and SLE}.---In the so-called Coulomb gas (CG) representation of the $O(n)$ model in the \textit{Euclidean} plane \cite{1982PhRvL..49.1062N}, one has the  parameterization $n=-2\cos \pi g$ with, for $n\in [0,2]$, a coupling constant $g\in [1/2,1)$ in the dense phase and $g\in [1,3/2]$ in the dilute phase. The critical scaling limit is known to correspond to Schramm-Loewner evolutions $\textrm{SLE}_\kappa$, or more precisely  to \textit{conformal loop ensembles} $\textrm{CLE}_\kappa$ \cite{springerlink:10.1007/s00220-009-0731-6,*2010arXiv1006.2373S}, with a parameter $\kappa=4/g$. 

 In terms of the critical exponent $a$ defined in \eqref{na}, we thus have the parameter relations  
\begin{equation}\label{ga}
g=a-1={4}/{\kappa}\in \left({1}/{2},{3}/{2}\right).
\end{equation} 

\textit{Quantum scaling exponents and KPZ}.---Given a fractal $\mathcal X$  on a  random surface $\mathcal S$,  its quantum fractal measure $\mathcal Q$ (i.e., the mass of $\mathcal X$) in a $\mathcal S$-subset  of area $\mathcal A$ scales as 
\begin{equation}\label{quantdim}
\mathcal Q\asymp \mathcal A^{1-\Delta},
\end{equation}
where $\Delta$ is the \textit{quantum scaling (KPZ) exponent} of $\mathcal X$. It is related to the  Euclidean scaling exponent $x$ of  $\mathcal X$, of fractal dimension $d=2-2x$, via  the celebrated KPZ relation \cite{MR947880,MR981529,*MR1005268}, rigorously proven in \cite{2009arXiv0901.0277D,*springerlink:10.1007/s00222-010-0308-1}:
\begin{equation}\label{kpz}
x=\gamma^2 \Delta^2/4+ \left(1-\gamma^2/4\right)\Delta,
\end{equation}
where $\gamma$ is the so-called \textit{quantum Liouville parameter} of the random surface. 

When the fractal $\mathcal X$ is part  of a geometrical critical model, such as the $O(n)$ loop model considered here, $\gamma$  depends on the model's central charge $c$  \cite{MR947880,MR981529,*MR1005268}, and in the CG-$\textrm{SLE}_\kappa$ framework (where $c=(6-\kappa)(6-16/\kappa)/4$) \cite{2010arXiv1012.4800D}:
\begin{equation}\label{gammakappa}\gamma^2=({4}/{g})\wedge 4g=\kappa \wedge ({16}/{\kappa}) \,\,\,(\leq 4).\end{equation}

\textit{Quantum Hausdorff dimensions in the dilute phase}.---We now claim the following relation between the Hausdorff dimension $D_\mathcal G$ of the  gasket $\mathcal G$ of a \textit{dilute} critical $O(n)$ model in quantum gravity, and that $D_\mathcal H$ of  
$\mathcal S$:
 \begin{equation}\label{basicscaling}
D_\mathcal G= D_\mathcal H(1-\Delta_\mathcal G),
\end{equation}
where $\Delta_\mathcal G$ is the gasket quantum scaling exponent. We explain this relation as follows. Denote by $\mathfrak l_{\mathcal G}$  the length of an  element of \textit{geodesic} in  $\mathcal G$. The gasket quantum measure  $\mathcal Q_\mathcal G$ in a ball of radius $\mathfrak l_{\mathcal G}$ in $\mathcal G$ is, with  \eqref{haus}, 
  \begin{equation}\label{Geodesic}
 \mathcal Q_\mathcal G \asymp \mathfrak l_{\mathcal G}^{\,D_\mathcal G}. 
 \end{equation} 
  A natural but crucial question is then \textit{whether an element of geodesic on the gasket $\mathcal G$ is also (locally) a geodesic of the whole fractal surface $\mathcal S$}. 
 We expect this to be true for the dilute phase of the $O(n)$ model, as explained now.
  
 In the Euclidean plane, the fractal dimension of the loop contact points is  $d_4=2-2x_4$, with a  ``four-leg'' exponent  $x_4=(3g+2-g^{-1})/4$. 
 Thus  $d_4\leq 0$ in the dilute phase $g\geq 1$. In quantum gravity,  the corresponding exponent  is obtained from $x_4$ by using KPZ  \eqref{kpz}, \eqref{gammakappa} for $g\geq 1$:  $\Delta_4=(1+g)/2 \geq 1$ \cite{1990NuPhB.340..491D,MR2112128}.  This implies a \textit{negative} quantum fractal dimension $D_4$ [$=D_\mathcal H(1-\Delta_4)$] for loop contact points.  In the scaling limit, they disappear and locally the quantum gasket should look like a  patch of random surface, with matching geodesics.
  
The quantum area $\mathcal A$ in the $\mathcal S$-ball of \textit{shared geodesic} radius $\mathfrak l=\mathfrak l_{\mathcal G}$ is then, by definition of  $D_\mathcal H$,  
\begin{equation}\label{geodesicS}\mathcal A\asymp \mathfrak l^{\,D_\mathcal H}.\end{equation} 
The Hausdorff dimension identity \eqref{basicscaling} then simply follows from comparing \eqref{Geodesic} and  \eqref{geodesicS} to \eqref{quantdim}.  (See also \cite{1996PhLB..388..713Ater}.) 

The  gasket Euclidean fractal dimension $d_\mathcal G$ was first determined 
by  CG techniques  \cite{PhysRevLett.63.2536,*PhysRevLett.64.493}, and recently rigorously derived  \cite{springerlink:10.1007/s00220-009-0731-6,*2010arXiv1006.2373S} as $d_{\mathcal G}=2-2 x_\mathcal G$, with 
$x_\mathcal G= (8-4g-3g^{-1})/16.$ 
The gasket quantum exponent 
is then obtained from  KPZ  \eqref{kpz}, \eqref{gammakappa} for $g\geq 1$:  
$\Delta_\mathcal G= (3-2g)/4.$ 
By combining this with \eqref{haus}, \eqref{ga} and \eqref{basicscaling}  we  get 
\begin{eqnarray}\label{four}
D_\mathcal H =4, 
\end{eqnarray}
independent of the value of $g \in [1,3/2)$, i.e., of the universality class of the  dilute critical $O(n)$ model.  Note that in that phase  $c\in [0,1]$ for $n\in [0,2]$, after inclusion of   
the pure gravity case $c=0, n=0$. 

Dilute critical lines have a  ``two-leg'' quantum exponent $\Delta_2=1/2$, independent of $n$, which governs the scaling of 
their quantum length 
$\mathfrak L\asymp {\mathcal A}^{1-\Delta_2} \asymp {\mathcal A}^{1/2}$ \cite{1990NuPhB.340..491D,MR2112128}. The gasket \textit{boundary}  exponent is given by \cite{MR2112128}: $\tilde \Delta _\mathcal G=2\Delta_\mathcal G-(1-g)=1/2=\Delta_2$, viz., the gasket's boundaries indeed are loops. From \eqref{four}, we thus conclude that  the quantum fractal dimension of dilute loops or gasket boundaries is $D_2=D{_\mathcal H}(1-\Delta_2) =2, \forall n \in [0,2]$.  

\textit{Quantum Hausdorff dimensions in the dense phase.}---This phase  corresponds  to $g\in (1/2,1)$, hence $c\in (-2,1)$.  The fractal dimension of the surface $\mathcal S$ is still $D_\mathcal H= 4$, but its derivation  requires a  modified geometrical argument. 

From KPZ \eqref{kpz}, \eqref{gammakappa}  applied to $x_4$ above for $g<1$, we get  a quantum contact exponent $\Delta^D_4=(3-g^{-1})/2<1$ in the dense phase \cite{1990NuPhB.340..491D,MR2112128}. Hence, like  the Euclidean one $d_4$, the quantum dimension $D_4 \propto (1-\Delta^D_4)$ for contact points is now positive and we have to take into account the non-simple,  infinitely bouncing, nature of dense loops. 
The relation between the loop quantum length and the area is  known to be modified to $\mathfrak L\asymp {\mathcal A}^{1-\Delta^D_2}\asymp  {\mathcal A}^{\frac{1}{2\nu}}$, with an exponent $\Delta_2^D=1-(2g)^{-1}$ and $\nu=g$ \cite{1990NuPhB.340..491D,MR2112128}.

We now distinguish  the length $\mathfrak l_\mathcal G$ of an element of geodesic on $\mathcal G$ from the (shorter) length $\mathfrak l$ of the corresponding direct geodesic on $\mathcal S$. In the scaling limit, the $\mathcal G$-geodesic, continually pinched at all scales by dense loops,  is  forced to pass through all their contact points.  It thus seems natural to assume that the same scaling relation holds between ${\mathfrak l}_{\mathcal G}$ and 
 ${\mathfrak l}$ as holds between $\mathfrak L$ and the typical length ${\mathcal A}^{1/2}$:
 \begin{eqnarray}\label{ll}
 {\mathfrak l}_{\mathcal G}\asymp {\mathfrak l}^{\,1/\nu}\asymp {\mathfrak l}^{\,1/g} \gg \mathfrak l. 
\end{eqnarray} 
Eqs. \eqref{Geodesic} and  \eqref{geodesicS} inserted into  
 relation \eqref{quantdim} then give
\begin{eqnarray}\label{basicscalingdense}
D_\mathcal G= \nu D_\mathcal H(1-\Delta^D_\mathcal G)
\end{eqnarray}
instead of \eqref{basicscaling}, and with $\Delta^D_{\mathcal G}$ the dense gasket quantum exponent. The latter  
is  obtained from KPZ  \eqref{kpz}, \eqref{gammakappa} for $g<1$, and from the same  $x_{\mathcal G}$ as above: 
$\Delta^D_\mathcal G= (2-g^{-1})/4.$ 
Using \eqref{haus}  in \eqref{basicscalingdense} again gives $D_{\mathcal H}=4, \forall g\in (1/2,1)$, thus $\forall c\in (-2,1)$. By continuity, this should also hold at $c=-2$ (and $1$).  
 The gasket has a boundary exponent \cite{MR2112128} $\tilde \Delta^D_\mathcal G=2\Delta^D_\mathcal G=\Delta^D_2$, viz., is made of loops  with a quantum fractal dimension in $\mathcal G$: $D_2=\nu D_{\mathcal H}(1-\Delta^D_2)=2.$ 
 
\textit{Concluding remarks.}---Recall that the Fortuin-Kasteleyn representation of the critical $Q$-state Potts model 
yields a  dense loop model with $n^2=Q \in [0,4]$, so that our result should also apply to the Potts model  in quantum gravity. 
The dilute and dense loop-model phases share the interval $c\in [0,1]$, and have the same dimension there. In particular, $c=0$ corresponds to  dilute $n=0$ self-avoiding walk or dense $n=1$ critical percolation models. Their Euclidean partition functions are  both $1$, which leads to  pure gravity  with $D_{\mathcal H}=4$. 

 For (non-unitary) $c<0$ models, numerical values  less than $4$  have been reported for $D_\mathcal H$ in \cite{PhysRevLett.68.2113,Ambj퓊n1998673,Kawamoto2002533}, in particular for   $c=-2$ (spanning trees), while recent simulations, by contrast, suggest there $1/D_{\mathcal H}\gtrsim 0.26$ \cite{Bernardi}. Admittedly, our result rests on assumption \eqref{ll}, but  for $g=1/2, c=-2$, this just agrees with the further anomalous boundary  scaling found in \cite{Ambj퓊n1998673}.   A direct combinatorial study of the geodesic scaling and of the superuniversality of the Hausdorff dimension would thus be especially interesting. 
 
 Lastly, the so-called $m$-multicritical models ($m\in \mathbb N$) have fractal dimensions $D_\mathcal H=2(m+1)$ \cite{Bouttier2003535}, with  $c \leq -22/5$ for $m\geq 2$, hence $c$ alone does not determine $D_\mathcal H$.    
 
\begin{acknowledgments}
Support by grant ANR-08-BLAN-0311-CSD5 is gratefully acknowledged.  I thank  O. Bernardi and G. Chapuy for  
communicating  unpublished results,  
E.  Guitter for useful discussions about Ref. \cite{2011arXiv1106.0153B}, and D. Kosower for a critical reading of the manuscript. 
\end{acknowledgments}
\bibliography{slekpz}
\bibliographystyle{apsrev4-1}\end{document}